%% file: schulte_Fuzz_2024_paper.tex
\documentclass[conference]{IEEEtran}
\IEEEoverridecommandlockouts
\usepackage{cite}
\usepackage{amsmath,amssymb,amsfonts}
\usepackage{accents}
\usepackage{algorithmic}
\usepackage{graphicx}
\usepackage{textcomp}
\usepackage{xcolor}

\def\BibTeX{{\rm B\kern-.05em{\sc i\kern-.025em b}\kern-.08em
    T\kern-.1667em\lower.7ex\hbox{E}\kern-.125emX}}
\begin{document}

\title{Coherent Design of Wind Turbine Controllers Considering Transitions between Operating Regions using Fuzzy Membership Functions}
	

\author{
	\IEEEauthorblockN{Horst Schulte}\\
	\IEEEauthorblockA{University of Applied Sciences Berlin (HTW Berlin) \\
		Faculty: School of Engineering - Energy and Information Science\\
		Control Systems Group, 12459 Berlin, Germany\\
		Email: schulte@htw-berlin.de}
}

\include{def_RT_v9}

\newcommand{\ubar}[1]{\underaccent{\bar}{#1}}
\def\kb{\Obs{k}}

\maketitle

\begin{abstract}  
	This paper presents a coherent design of wind turbine controllers with explicit consideration of transitions between operating regions by fuzzy membership functions. In improving the design process of wind turbines, the transitions between partial-load operation by torque control and full-load operation by pitch control need to be systematically considered. From the first view, fuzzy methods for blending separately designed control laws are an obvious choice. However, valid design rules must be developed to ensure stability and performance during the transition. A model-based control design procedure in the Takagi-Sugeno fuzzy framework using the sector nonlinearity method is proposed to achieve the above control design objectives. In addition to a detailed mathematical analysis of the design, the method's applicability is verified by simulation studies using a high-fidelity reference wind turbine model. 

\end{abstract}


This work has been submitted to the IEEE for possible publication. Copyright may be transferred without notice, after which this version may no longer be accessible.

\section{Introduction}
\noindent The importance of wind turbines in the generation mix of electrical energy is growing steadily \cite{Komarnicki2023}. As a result, the demands on the reliability and performance of wind turbines are also increasing, as well as the professionalization of engineering in this area \cite{wes-1-1-2016}. Science policy accounts for this with funding of research projects and academic education programs.  As a result of funding strategies, among other things, four main trends can be identified in wind turbine research and development from a system and control theory perspective.\\ The first two research directions, the {\it Load Mitigation Control} (MTC) and {\it Fault Tolerant Control} (FTC), are related to the advanced control of single wind turbines. Based on the control objectives of optimized power generation in the partial-load and limited power generation in the full-load region,
the control design methods for wind turbines are extended to include the control objectives of load mitigation and fault-tolerant control under sensor and actuator faults. The other two research areas, {\it Wind Farm Control} (WFC) and {\it Wind Power Integration} (WPI) are dedicated to wind turbines as part of a wind farm and their integration into the electrical grid.\\
An initial approach for a MTC scheme for structural load reduction is based on individual pitch control. Here, the motion of three wind turbine blades is divided into three collective eigenmotions using the Coleman transformation to systematically design a decoupled scheme by controlling the pitch of each blade independently \cite{Bossanyi2003}. That so-called individual pitch control (IPC) is superimposed on the standard collective pitch control signal in the full-load region. Another concept of the MTC utilizes a feedforward coupling in the pitch controller whereby the measured wind speed in front of the rotor \cite{SchlipfCheng2013} or a wind speed observer is used \cite{GauterinKammerer.2015} to adjust the pitch angle. This allows the pitch angle to be changed without first activating the pitch angle via the rotor speed control error. With the appropriate setup, the loads on the tower can be reduced. More recent studies, i.e., presented in \cite{Bossanyi2003} utilize control methods such as the $H_\infty$ method for robust IPC control \cite{KipchirchirSoeffker} or LIDAR-assisted feedforward individual pitch control \cite{RussellColluMcDonaldEtAl}. The MTC concepts mentioned are only active in the full-load region. A transition between the generator control (partial-load region) and pitch control (full-load region) is not considered. In practice, however, this is important because the turbine control in the transition must be well-designed to prevent bumps at the controller output.\\
FTC of wind turbines aims to increase the availability of wind turbines integrated into electrical grids and to improve the reliability of island grids with wind turbines. One successfully developed FTC concept for wind turbines is the fault hiding approach
\cite{BlankeKinnaert.2006}. For this purpose, a reconfiguration block is connected between the controlled system, actuator(s) and sensor(s), and the nominal power tracking controller. The reconfiguration block must meet the requirement that the behavior of the reconfigured system corresponds as closely as possible to the behavior of the nominal, i.e., fault-free system without sensor and actuator faults. A comparison of many different FTC approaches was carried out in \cite{Odgaard_FDI_results:2012} using a wind turbine benchmark model \cite{Odgaard:2009} with sensor and actuator faults \cite{Odgaard:FAST_benchmark}. Typically, the proposed FTC schemes are extensions of baseline controllers or disturbance observer-based state feedback designed for individual wind turbine regions.\\ 
Wind farm controllers aim is to optimize the overall power generation and reduce the mechanical loads on each turbine. In the last decade, new control concepts for the overall control of wind farms have been proposed in the literature \cite{KnudsenBakSvenstrup2015}. Either aerodynamic or electrical optimization aspects \cite{NguyenKim2021} are the focus of the research. Due to the flow conditions in a wind farm and the complex interaction effect of the mechanical structure with the airflow, not all aspects of the research direction have been investigated yet \cite{StockLeithead2022}. To the author's knowledge, recent efforts to reduce the loads on individual turbines through higher-level wind farm design \cite{AnderssonAnaya-LaraTandeEtAl2021} have not considered the transitions between region controllers. Usually, simulations are performed above or below the nominal wind speed. Transitions are not explicitly taken into account, and their effect on the mechanical loads and lifetime of a turbine is not analyzed.\\
However, there is a reference in the textbook \cite{Burton2021} and a research paper \cite{PoeschkeGauterinKuehnEtAl2020} on how to design the transition between torque control in the partial-load and pitch control in the full-load region. In \cite{Burton2021} (Chapter~8), a smooth bumpless transition is done by independently adjusting the reference signal of the torque and pitch control. It allows the integration of hysteresis in the transition without changing the closed-loop structure or control parameters.  
The disadvantage is the lack of a theorem-based mathematical design. On the other hand, in \cite{PoeschkeGauterinKuehnEtAl2020}, a systematic design is proposed in the Takagi-Sugeno fuzzy framework using weighted combinations of integral state feedback matrices. The disadvantage of that method is the large number of local models required, which are determined for each region using Taylor linearization to represent the dynamics of the plant approximately. The transition is performed using weight matrices that make the transitions bumpless. In contrast to the previously mentioned local TS approach \cite{PoeschkeGauterinKuehnEtAl2020} 
and described heuristic in \cite{Burton2021}, this paper proposes a well-defined global approach with nonlinear sector functions as membership functions and freely adjustable fuzzy membership functions for the transition with fewer tuning parameters.\\  
The structure of the paper is as follows: Section~\ref{sec:Problem_Formulation} first presents the essentials of single wind turbine control systems, focusing on primary power conversion (from wind to generator power). This is followed by describing the region controllers for power optimization and limitation. A brief description of the system and control structure are given based on results presented in \cite{PoeschkeSchulte2021}. The main contribution of the paper is given in Section~\ref{sec:fuzzy_blending}. Here, a convex combination of individual controllers from the previous section is introduced with the purpose of integration into the LMI-based optimization framework. In Section~\ref{sec:simulation_results_discussion}, the simulation results for a 5~MW reference wind turbine are presented and discussed. The paper ends with the conclusion in Section~\ref{sec:conclusion}.

\section{Problem Formulation}
\label{sec:Problem_Formulation}

\subsection{Essentials of State of the Art Wind Turbine Control}
\label{sec:EssentialsStateOfArt}
\begin{figure}[ht]
	\centering
	\includegraphics[width=0.45\textwidth]{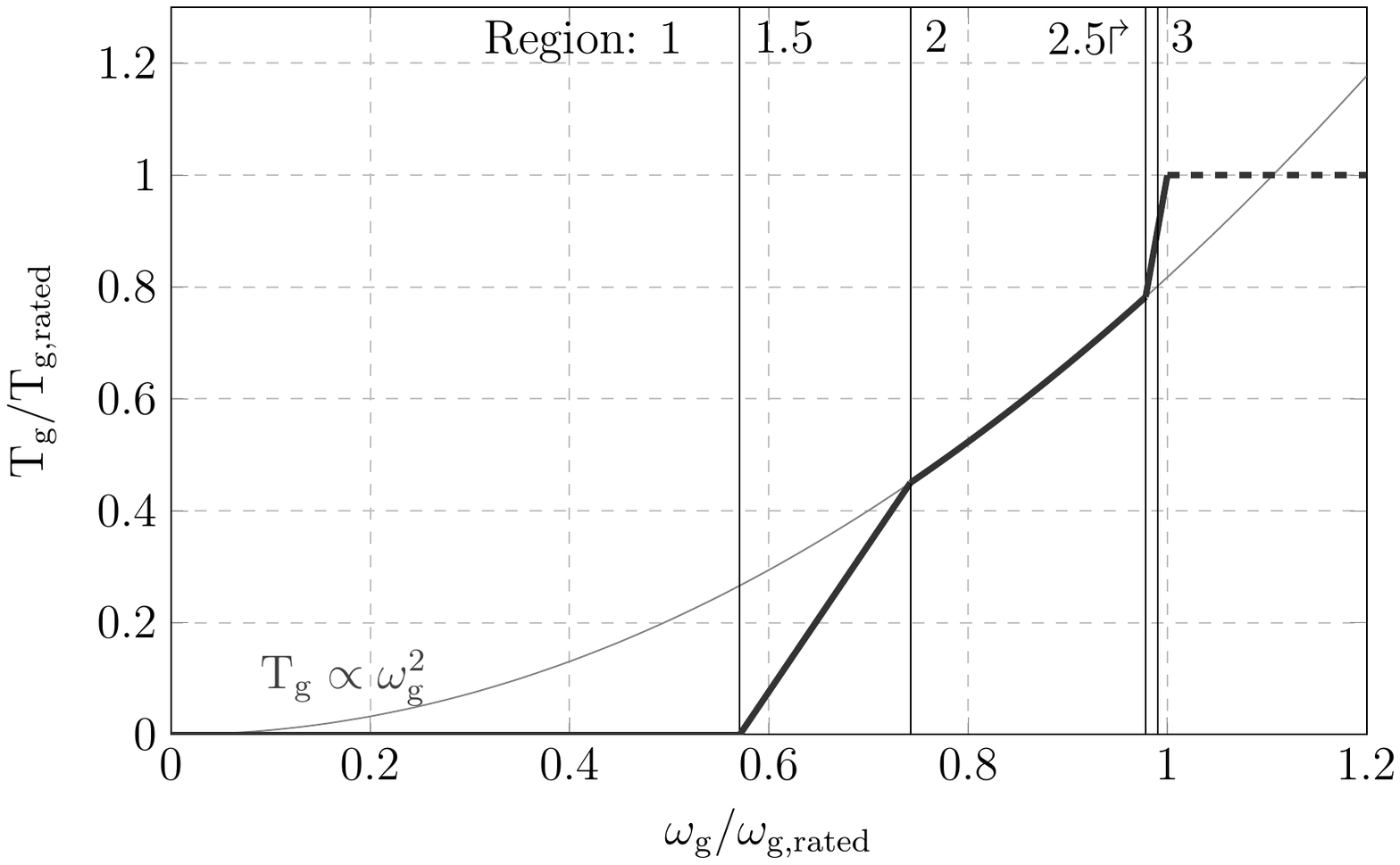}
	\caption{$T_g$ as a function of the generator speed $\omega_g$}
	\label{fig:WTC:OperationOverall}
\end{figure}

\noindent For a better understanding of the proposed fuzzy blending concept in Section~\ref{sec:fuzzy_blending}, the essentials of wind turbine control are explained briefly. The operation of wind turbines is broadly divided into two main regions. Below the rated wind speed $v_{rated}$, the rotor speed $\omega_r$ is less than the rated rotor speed. This region is called the partial-load region because the turbine generator power is less than the rated power. Therefore, the main control objective in the partial-load region is power optimization. If the wind speed exceeds the rated wind speed, the generator power is actively limited to the rated power. This region is called the full-load region. The partial load is divided into the sub-regions 1.5, 2, and 2.5 illustrated in Figure~~\ref{fig:WTC:OperationOverall}. In these sub-regions, control takes place by adjusting the generator torque $T_g$ to counteract the aerodynamic torque of the turbine rotor $T_r$. The generator torque is either open-loop controlled as a function of the generator speed or set by a wind speed observer-based tracking controller. With the latter concept, the wind turbine is more precisely controlled in the closed-loop. This will be presented in Section~\ref{sec:SFC_in_partial_load}.\\ 
But first to the baseline concept. The associated generator curve is shown in Figure~\ref{fig:WTC:OperationOverall}. The generator torque to reach the optimal power generation is proportional to the square of the generator speed but is only set in Region~2, i.e., if the wind speed $v$ is sufficiently high ($v > v_{\text{cut,in}}$) but below the rated generator speed $\omega_{\text{g,rated}}$ ($\omega_{\text{g}} < \omega_{\text{g,rated}}$). 
In Region~3, the full-load region, the generator torque is usually set to a constant torque, the rated torque $T_{\text{g,rated}}$, see Figure~\ref{fig:WTC:OperationOverall}. To keep the turbine within the permissible speed range at wind speeds above the nominal speed, the rotor blades are pitched to reduce the aerodynamic torque factor so that the rotor torque does not exceed the nominal generator torque at the rotor shaft.\\ 
The automatic decision uses transition conditions to decide which region controller is active. These conditions use the generator speed and torque sensing. 
Section~\ref{sec:fuzzy_blending} describes how to formalize the transitions between two multi-output control laws in a model-based fuzzy framework. Firstly, we introduce the two region controllers in Section~\ref{sec:SFC_in_partial_load} and Section~\ref{sec:SFC_in_full_load}. These are then connected to form a coherent control law without switching heuristics by convex fuzzy sets in Section~\ref{sec:fuzzy_blending}.

\subsection{Control scheme in the Partial-load Region}
\label{sec:SFC_in_partial_load}
\begin{figure}[ht]
	\centering
	\includegraphics[width=0.42\textwidth]{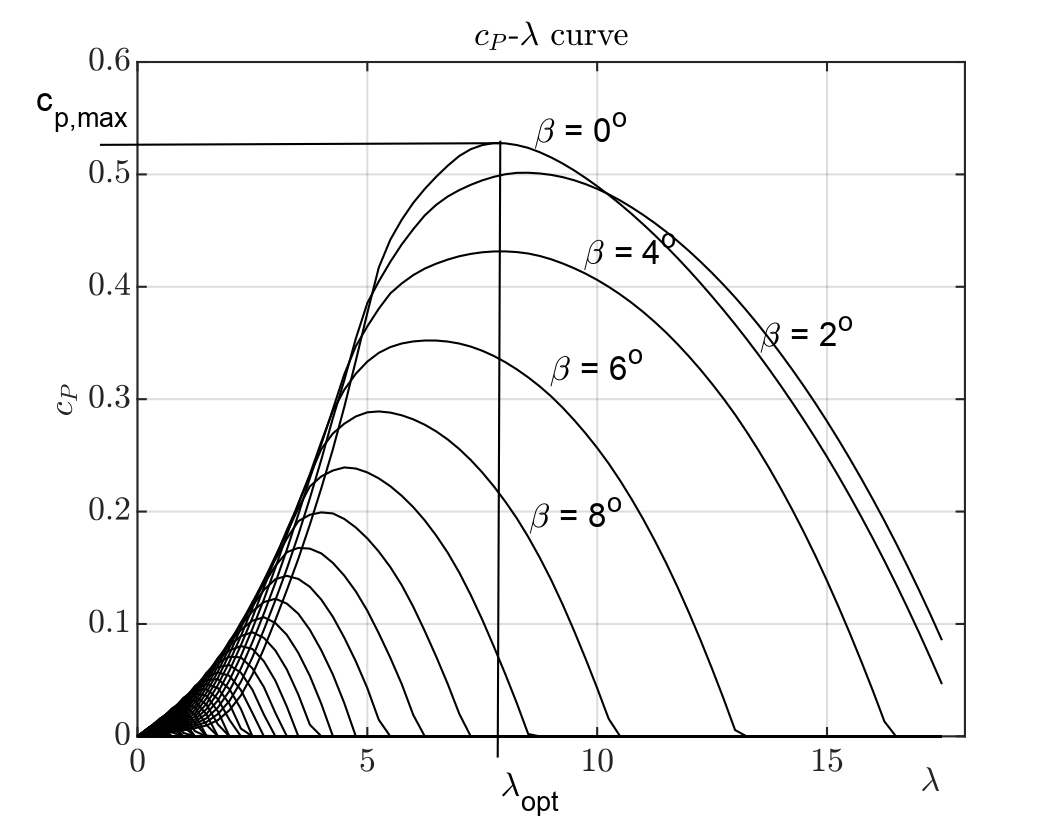}
	\caption{$c_P$-$\lambda$ curve with labeled $c_{P,max}$ point and related $\lambda_{opt}$ value}
	\label{fig:cP_lambda_curve}
\end{figure}

\begin{figure}[ht]
	\centering
	\includegraphics[width=0.48\textwidth]{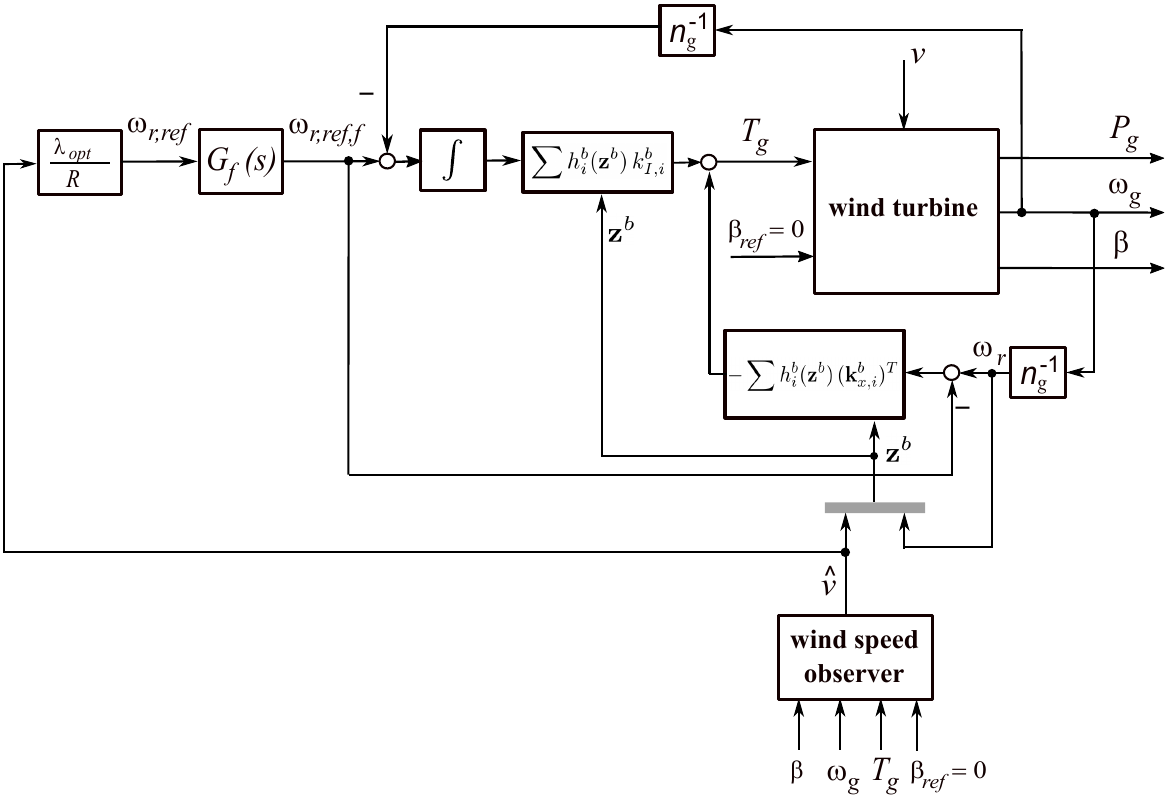}
	\caption{Partial-load region control scheme with integral state-feedback PDC model-based TS fuzzy control law \eqref{eq:cntrl_law_R2} and TS wind speed observer  \cite{GauterinKammerer.2015}}
	\label{fig:SFB_Control_Partial_Load}
\end{figure}

\noindent The control objective in the partial-load region is power optimization. This means that the power factor 
\begin{equation}
  \label{eq:def_cP}
  c_P = \frac{P_g(T_g,\omega_g)}{P_w(v)}
\end{equation}
defined by the ratio of generated wind turbine power $P_g = T_g \, \omega_g$ over the wind power $P_w(v) \propto v^3$ must be optimized. In \eqref{eq:def_cP} $\omega_g$ denotes the generator speed and $v$ the upstream wind speed in front of the rotor. The $c_P$-$\lambda$ curve represents the $c_P$ factor as a function of the dimensionless tip speed ratio  
\begin{equation}
  \label{eq:lambda}
  \lambda = \frac{\omega_r \,R}{v}
\end{equation}
and the pitch angle $\beta$ shown in Figure~\ref{fig:cP_lambda_curve}. The pitch angle\footnote{The rotor blades, usually three, are adjusted synchronously. Individual rotor blade pitch control (IPC) is not used.} as the rotation angle around the longitudinal axis changes the inflow at the blade. For the pitch angle $\beta = 0$, the wind turbine reaches its maximum power output with $c_{P,max}$ at the so-called $\lambda_{opt}$ point, see Figure~\ref{fig:cP_lambda_curve}. In Eq.~\eqref{eq:lambda}, the variable $\omega_r$ denotes the rotor speed, and the parameter $R$ is the rotor radius. A stiff drive train with $\omega_g =  n_g \, \omega_r$ is assumed with $n_g$ as the gear ratio.\\ 
To optimize the turbine, the controller adjusts the generator torque $T_g$ to regulate the rotor speed so that the maximum value of $c_P$ is tracked during changes in wind speed. The reference rotor speed is derived from the measurement of the wind speed using a LIDAR system \cite{SimleyPaoFrehlichEtAl2014} or by a model-based observer that reconstructs the wind speed \cite{GauterinKammerer.2015}:\begin{equation}
  \label{eq:omega_r_ref}
   \omega_{r,ref} = \frac{\lambda_{opt}}{R} \, v \, , \qquad v \in \{v_m, \hat{v} \}
\end{equation}
where $v_m$ denotes the measured and $\hat{v}$ the estimated wind speed. The control scheme of the partial-load region is illustrated in Figure~\ref{fig:SFB_Control_Partial_Load}.
The wind turbine has two control inputs: the reference pitch angle $\beta_{ref}$ for a lower-level pitch drive control and the generator torque. The latter is set directly due to disregarding the fast dynamics of the lower-level torque control of the generator. As shown in Figure~\ref{fig:SFB_Control_Partial_Load}, the reference pitch angle is zero in the partial-load operating region.
The controller comprises two PDC blocks for state feedback and controller error integration. In addition, a Takagi-Sugeno disturbance observer is used for wind speed estimation \cite{GauterinKammerer.2015}. To adjust the tracking signal dynamics, a filter is used 
\begin{equation}
    \omega_{ref,f}(s) = G_f(s) \, \omega_{ref}(s) \; .   
\end{equation}
Furthermore, since in typical wind turbines, the faster generator speed is measured instead of the slower rotor speed (60 to 220 times faster), the control scheme also contains the speed calculation to the rotor speed with $\omega_r = n_g^{-1} \, \omega_r $.

\subsection{Control scheme in the Full-load Region}
\label{sec:SFC_in_full_load}
\begin{figure}[ht]
	\centering
	\includegraphics[width=0.48\textwidth]{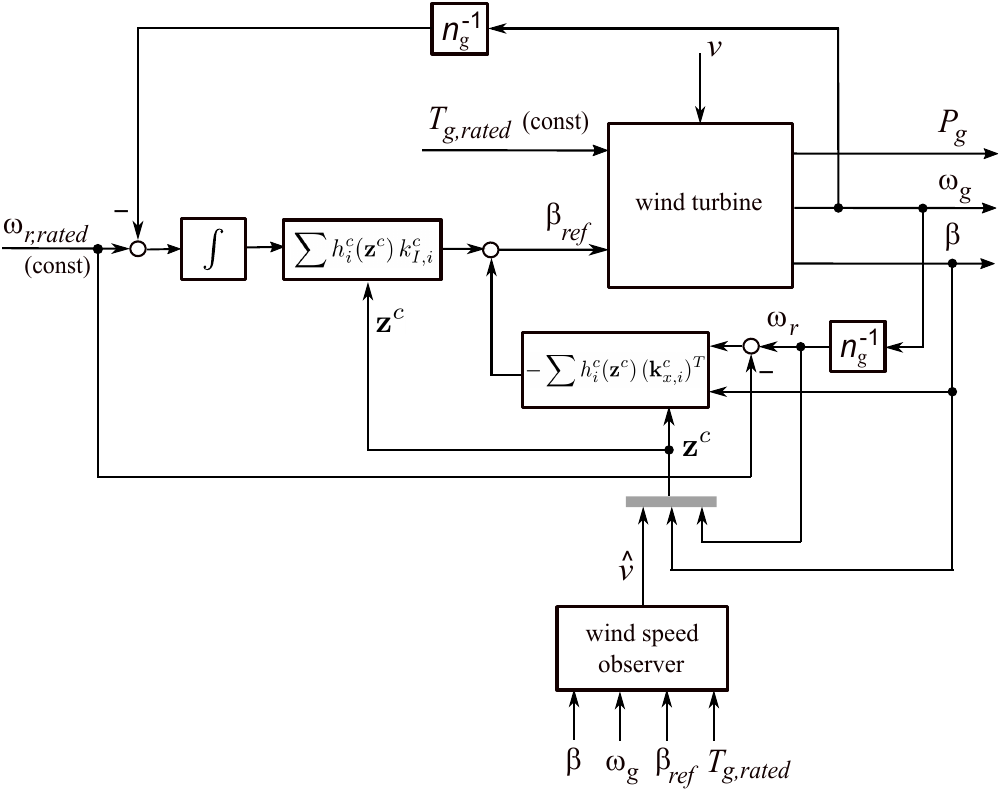}
	\caption{Full-load region control scheme with integral state-feedback PDC model-based TS fuzzy control law \eqref{eq:cntrl_law_R3} and TS wind speed observer  \cite{GauterinKammerer.2015}}
	\label{fig:SFB_Control_Full_Load}
\end{figure}

\noindent The control scheme in the full-load region is shown in Figure~\ref{fig:SFB_Control_Full_Load}. 
A significant difference from the previous scheme is that the pitch angle is the control variable, and the torque of the generator is set to the fixed nominal torque. As with the partial-load governor, the turbine speed is regulated but to a setpoint and not to a variable speed. To achieve the rated power, the turbine is kept at the rated speed with $\omega_{r,ref} = \omega_{r,rated}$ for wind speeds above the rated speed ($v  \ge v_{rated}$), regardless of the change in the wind speed. Note, if necessary, the turbine can operate permanently in deloading mode, e.g., to reduce noise emissions at night by reducing the setpoint $\omega_{r,ref} < \omega_{r,rated}$. However, due to the high inertia of the rotor, this approach is not suitable for rapid active power changes below the currently available power to support fast grid frequency control \cite{PoeschkePetrovicBergerEtAl2022}. 


\section{Fuzzy-Blending of Control Regions}
\label{sec:fuzzy_blending}

\begin{figure}[thpb!]
	\begin{center}
		\includegraphics[width=0.35\textwidth]{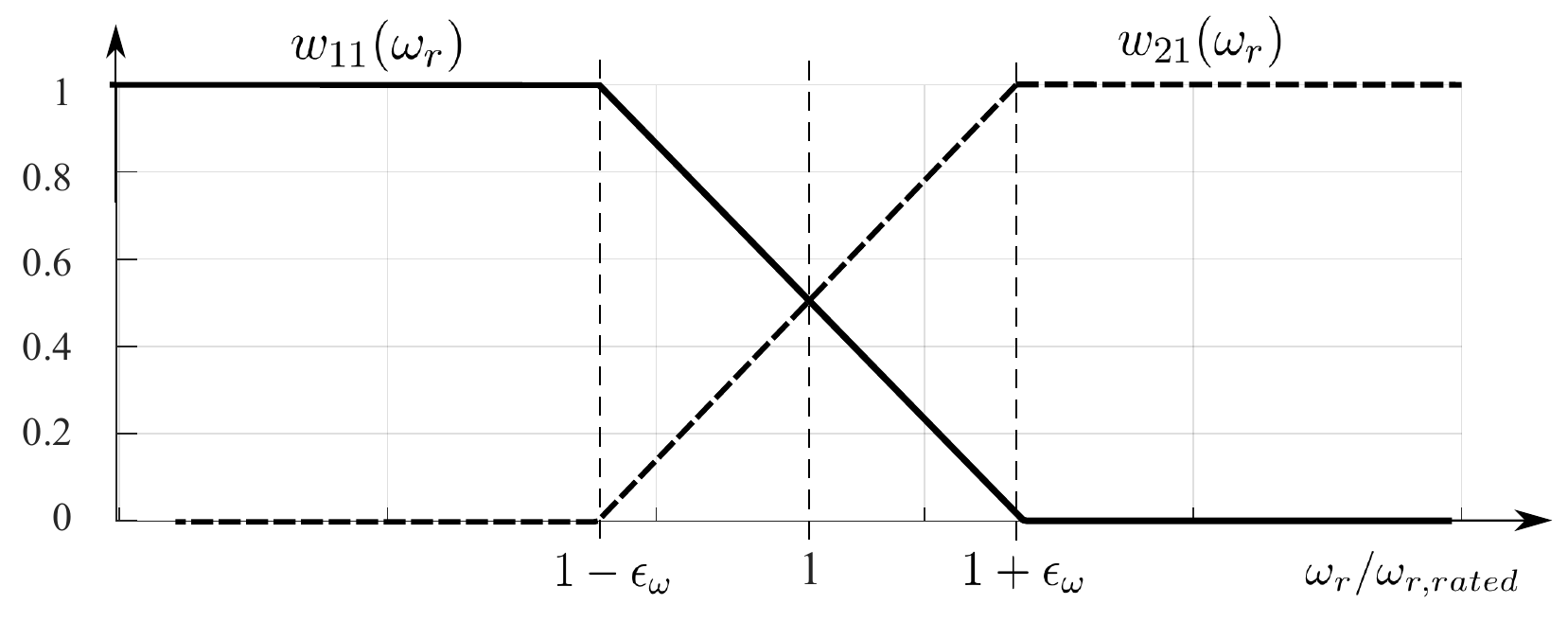} 
		\caption{Weighting functions $w_{i1}(\omega_r)$ with the overlapping factor  $\epsilon_\omega$}
		\label{Fig:weighting_functions_4_omega_r} 
	\end{center}
\end{figure}

\begin{figure}[thpb!]
	\begin{center}
		\includegraphics[width=0.35\textwidth]{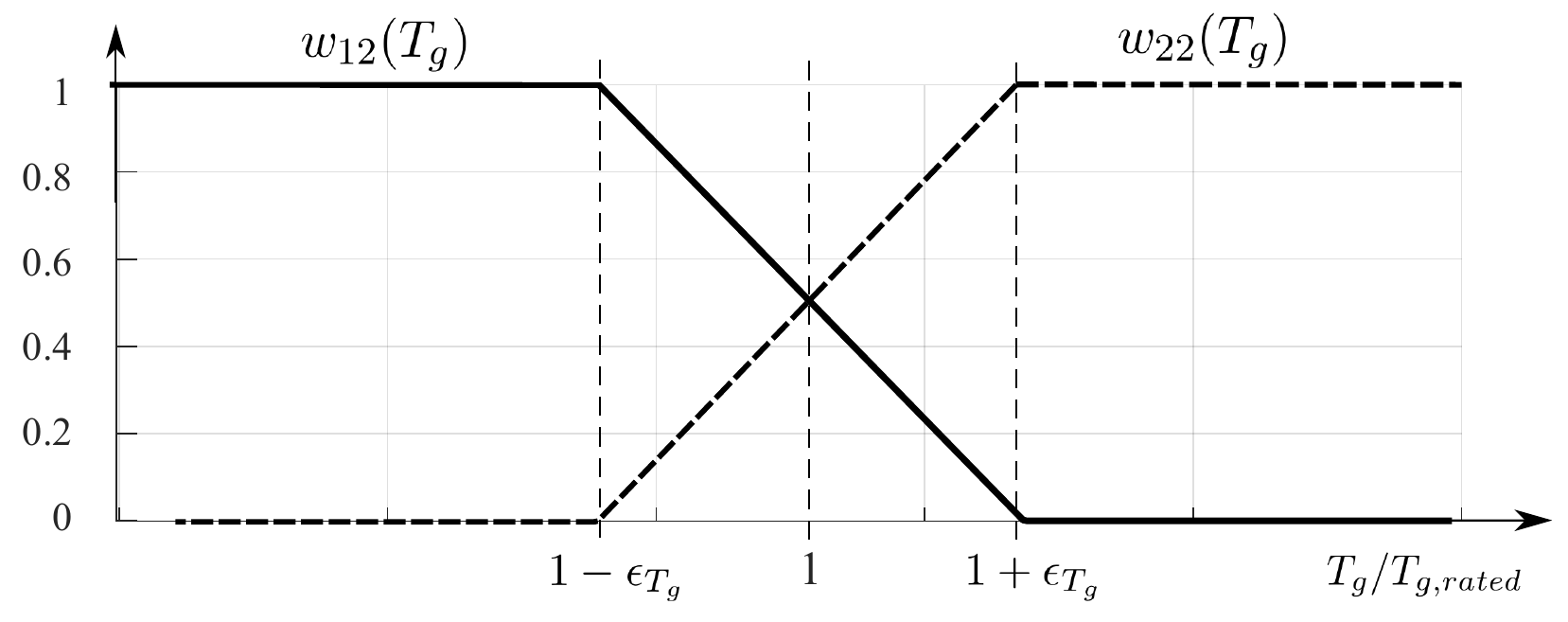} 
		\caption{Weighting functions $w_{i2}(T_g)$ with the overlapping factor  $\epsilon_{T_g}$}
		\label{Fig:weighting_functions_4_Tg} 
	\end{center}
\end{figure}

For a bumpless operation between partial-load and full-load, the control laws of these regions are merged using the fuzzy approach. For this purpose, a coupling filter for the transition 
is introduced first
{\small
\begin{align}
   \label{eq:bumpless_filter}
    \begin{pmatrix}
         \tilde\beta_{ref} \\
         \tilde{T}_g
     \end{pmatrix} = 
     \begin{pmatrix}
          f_{1} (\omega_r) & 0\\
              0 & f_{2}(T_g)  \\
     \end{pmatrix}
     \begin{pmatrix}
            \beta_{ref} \\
             {T}_g
      \end{pmatrix} = 
       \Fb(\omega_r, T_g) 
       \begin{pmatrix}
           \beta_{ref} \\
                {T}_g
           \end{pmatrix} ,
\end{align}}
where 
\begin{align*}
   f_{1} (\omega_r)  & = w_{11}(\omega_r) \, \ubar{f}_1  +  w_{21}(\omega_r) \, \bar{f}_1 \, , \\
   f_{2} (T_g)   & = T_{g,max} \, \frac{1}{T_g}  = w_{12}(T_g) \, \ubar{f}_2  +  w_{22}(T_g) \, \bar{f}_2  
\end{align*}
with the sector function $w_{ij}$ related to the lower and upper sector bounds
\begin{align}
   \ubar{f}_1 = 0 , \quad \bar{f}_1 = 1 , \quad
   \ubar{f}_2 = \frac{T_{g,max}}{T_{g,max}} = 1, \quad \bar{f}_2 = \frac{T_{g,max}}{T_{g,min}} \, .
\end{align}	
Due to the convex sum condition $w_{1j}(\cdot) + w_{2j}(\cdot) = 1$ illustrated in Figure~\ref{Fig:weighting_functions_4_omega_r} and Figure~\ref{Fig:weighting_functions_4_Tg}, the filter matrix \eqref{eq:bumpless_filter} can be written as 
\begin{align}
	\Fb(\omega_r, T_g)  = 
	 \begin{pmatrix}
	  F_{11} & 0\\
	  0 & F_{22}\\
	\end{pmatrix}
\end{align}
with 
\begin{align*}
   F_{11} & = \big(w_{12}(T_g)  +  w_{22}(T_g) \big) \big(w_{11}(\omega_r) \, \ubar{f}_1  +  w_{21}(\omega_r) \, \bar{f}_1\big)\\
   F_{22} & = \big(w_{11}(\omega_r) +  w_{21}(\omega_r) \big) \big( w_{12}(T_g) \, \ubar{f}_2  +  w_{22}(T_g)\bar{f}_2   \big)   
\end{align*}
and rearranged in
\begin{align*}
\Fb(\omega_r, T_g)  = 
& \underbrace{w_{11}(\omega_r) w_{12}(T_g)}_{h^a_1(\omega_r, T_g)} 
\underbrace{\begin{pmatrix}
	\ubar{f}_1  & 0\\
	0      &   \ubar{f}_2\\
	\end{pmatrix}}_{\Fb_1}\\ 
& +
\underbrace{w_{11}(\omega_r) w_{22}(T_g)}_{h^a_2(\omega_r, T_g)} 
\underbrace{\begin{pmatrix}
	\ubar{f}_1  & 0\\
	0      &   \bar{f}_2\\
	\end{pmatrix}}_{\Fb_2} \\
& + \underbrace{w_{21}(\omega_r) w_{12}(T_g)}_{h^a_3(\omega_r, T_g)}  
\underbrace{\begin{pmatrix}
	\bar{f}_1  & 0\\
	0      &   \ubar{f}_2\\
	\end{pmatrix}}_{\Fb_3} \\ 
& +
\underbrace{w_{21}(\omega_r) w_{22}(T_g)}_{h^a_4(\omega_r, T_g)}   
\underbrace{\begin{pmatrix}
	\bar{f}_1  & 0\\
	0      &   \bar{f}_2\\
	\end{pmatrix}}_{\Fb_4}   \, ,
\end{align*}
which gives the compact sum form
\begin{align}
  \label{eq:Bumpless_Filter} 
  \Fb(\omega_r, T_g) = \sum_{i=1}^{N^a_r=4} h^a_i(\zb^a) \, \Fb_i 
\end{align}
with $\zb^a = (\, \omega_r \, , \, T_g \,)^T$. 
The state controller shown as a block structure in Figure~\ref{fig:SFB_Control_Partial_Load} is given as
\begin{align}
	\label{eq:cntrl_law_R2}
	u_1 = -\sum_{i=1}^{N^b_r} h^b_i(\zb^b) \, \underbrace{(k^b_{x,i} \, , \, 0)}_{(\kb^b_{x,i})^T} \, \Delta{\xb} +  \sum_{i=1}^{N^b_r} h^b_i(\zb^b) \, k^b_{I,i} \, x_I  \; ,   
\end{align}
where $u_1 = T_g$ with $\zb^b = (\, \omega_r \, , \, \hat{v} \,)^T$. And the control law of the full-load region from Figure~\ref{fig:SFB_Control_Full_Load} with $u_2 = \beta$ as controller output is 
\begin{align}
  \label{eq:cntrl_law_R3}
  u_2 = -\sum_{i=1}^{N^c_r} h^c_i(\zb^c) \, (\kb^c_{x,i})^T \, \Delta{\xb} +  \sum_{i=1}^{N^c_r} h^c_i(\zb^c) \, k^c_{I,i} \, x_I   \; ,
\end{align}
where $\zb^c = (\, \omega_r \, , \, \beta \, , \, \hat{v} \,)^T$.
Note the superscripts $x~\in~\{a,b,c\}$ are introduced to distinguish the membership functions, premise variables, and controller gains of the two regions and the decoupling filter. The two control laws \eqref{eq:cntrl_law_R2} and \eqref{eq:cntrl_law_R3} can be related to the common state vector \begin{align}
    \Delta\xb = 
    \begin{pmatrix}
        \Delta x_1\\
        \Delta x_2
     \end{pmatrix}
    = 
    \begin{pmatrix}
        \omega_r - \omega_{r,ref}\\
        \beta 
    \end{pmatrix}
\end{align}
and auxiliary integral state 
\begin{align}
  x_I := \int_{0}^{t} (\omega_{r,ref} - \omega_r) \,  d\tau =
  \int_{0}^{t} e \, d\tau   \; .
\end{align}
Even if the first control law does not change the pitch angle, it can be used for an aggregated control law given as follows 
\begin{align}
  \begin{split}
   \label{eq:common_control_law}
   \begin{pmatrix}
      u_1\\
      u_2
    \end{pmatrix} =
     - \sum_{i=1}^{N^b_r} \sum_{j=1}^{N^c_r}  & h^b_i(\zb^b)  h^c_j(\zb^c) \\ 
  & \cdot  \begin{bmatrix}
    \begin{pmatrix}
      (k^b_{x,i} \, , \, 0) \\
      (\kb^c_{x,j})^T
     \end{pmatrix} \Delta{\xb} + 
       \begin{pmatrix}
         k^b_{I,i} \\
         k^c_{I,j}
       \end{pmatrix} {x_I} 
     \end{bmatrix}
    \end{split} 
\end{align}
and combined with the coupling filter \eqref{eq:Bumpless_Filter} the multi-region TS fuzzy control law is obtained 
\begin{align}
	\label{sec:multi-region_controller}
 \begin{split}
   \begin{pmatrix}
     \tilde{u}_1\\
     \tilde{u}_2
   \end{pmatrix} =
      - \sum_{m=1}^{N^a_r} & \sum_{i=1}^{N^b_r} \sum_{j=1}^{N^c_r}   h^a_m(\zb^a) h^b_i(\zb^b)  h^c_j(\zb^c) \\
    & \cdot \Fb_m
    \begin{bmatrix}
      \begin{pmatrix}
        (k^b_{x,i} \, , \, 0) \\
        (\kb^c_{x,j})^T
      \end{pmatrix} \Delta{\xb} + 
    \begin{pmatrix}
        k^b_{I,i} \\
        k^c_{I,j}
    \end{pmatrix} {x_I} 
  \end{bmatrix} \, .
\end{split}
\end{align}

\section{Simulation Results and Discussion}
\label{sec:simulation_results_discussion}
\begin{figure}[ht]
	\centering
	\includegraphics[width=0.39\textwidth]{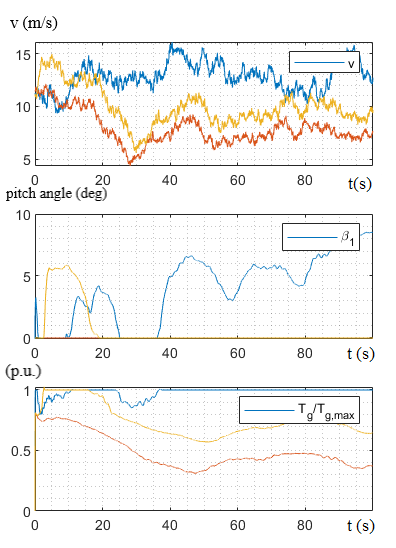}
	\caption{(a) v: Turbulent wind speed in front of the rotor with the same degree of turbulence but different mean values, (b) Pitch angle: Pitch angle adjustment by the multi-region controller \eqref{sec:multi-region_controller} in relation to three different wind speed curves illustrated by the pitch angle of rotor blade~1 denoted with $\beta_1$, (c) Generator torque $T_g$ in (p.u.): Generator torque adjustment of the multi-region controller \eqref{sec:multi-region_controller} with the three different wind speed curves}
	\label{fig:sim_v_pitch_Tg}
\end{figure}

\begin{figure}[ht]
	\centering
	\includegraphics[width=0.39\textwidth]{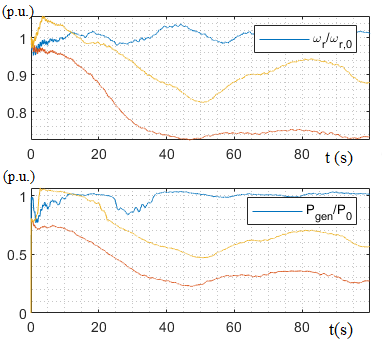}
	\caption{(a) Rotor speed $\omega_r$ in (p.u.): Rotor speed curve for the three different wind speeds given in Figure~\ref{fig:sim_v_pitch_Tg}, (b) Power generation in (p.u) for the three different wind speeds given in  Figure~\ref{fig:sim_v_pitch_Tg}}
	\label{fig:sim_omegar_Pg}
\end{figure}

\begin{figure}[ht]
	\centering
	\includegraphics[width=0.39\textwidth]{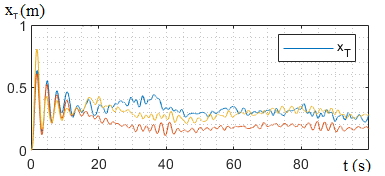}
	\caption{Wind turbine tower tip motion $x_T$ at in (p.u.) for the three different wind speeds given in Figure~\ref{fig:sim_v_pitch_Tg}}
	\label{fig:sim_xT}
\end{figure}

The simulation results for a 5~MW reference offshore wind turbine \cite{Jonkman:2009}, inspired by the Repower 5M turbine, with the multi-region controller \eqref{sec:multi-region_controller} calculated with the FAST simulation tool \cite{Jonkman2020} integrated in Simulink$\copyright$ are shown in Figure~\ref{fig:sim_v_pitch_Tg}, Figure~\ref{fig:sim_omegar_Pg} and Figure~\ref{fig:sim_xT}. The wind speeds are chosen so that the controller operates two resp. three times in the transition between Region~2 and Region~3, see Figure~\ref{fig:WTC:OperationOverall}. At the wind speed with the highest mean value (curve diagram in blue), the turbine controller operates for the most part in Region~3 and only in the time intervals of $t = [1, 9]$ sec and $t = [25, 35]$ sec in Region~2. This can be seen by comparing the pitch angle with the generator torque adjustment and recognized by comparing the pitch angle (blue curve) with the torque setting (blue curve) of the generator. A small overlap of the two manipulated variables is clearly visible. This is due to the corresponding overlap of the fuzzy sets in Figure~\ref{Fig:weighting_functions_4_omega_r} and Figure~\ref{Fig:weighting_functions_4_Tg} of the coupling filter \eqref{eq:bumpless_filter}.\\
Note that the legend in Figure~\ref{fig:sim_v_pitch_Tg} uses $\beta_1$ to denote the measured angle of blade~1. Due to the collective pitch adjustment, the pitch angles of the rotor blades are identical. Furthermore, due to the fast dynamics of lower-level control for blade angle adjustment, it is $\beta_{ref} \approx \beta_i$ for $x = 1,2,3$. 
In addition, a wind speed (red line in Figure~\ref{fig:sim_v_pitch_Tg}) at which the wind turbine is fully in the partial load range is also shown for a better understanding of the overall dynamics.\\ 
As an indicator of the effect of the smooth bumpless transition, the states of the turbine, the rotor speed, tower deflection, and generator power are also shown. It can be seen in
Figure~\ref{fig:sim_omegar_Pg} and Figure~\ref{fig:sim_xT} the region transition does not affect the turbine states. This behavior is desirable because it prevents additional vibrations in the mechanical components of a wind turbine, thus reducing the fatigue loads. This is a major benefit for offshore turbines in particular, as the large fluctuations in wind speed mean that changes between partial and full-load operation occur more frequently.\\ 
For a more precise evaluation of the influence of the degree of overlap $\epsilon_\omega$ in Figure~\ref{Fig:weighting_functions_4_omega_r} and Figure~\ref{Fig:weighting_functions_4_Tg}, investigations using the damage equivalent load (DEL) method would still have to be carried out. This method was used to improve the evaluation of wind turbine governors in \cite{Clemens2020} and will be used here in a further study to evaluate the impact of the new coherent controller design.

\section{Conclusion}
\label{sec:conclusion}
It is shown that the proposed method can be integrated into an existing TS controller design. Using the example of a wind turbine control system, where the dual-input system is influenced by only one input per controller region, two independently designed single-output control laws have been transferred to a single coherent dual-output controller by introducing a fuzzy coupling filter. Here, the partial and full-load controllers are designed separately and combined with the proposed fuzzy blending. Due to the consistent use of the TS convex sum notation, both for the local region controllers and the multi-region control law, stability proofs and a one-step synthesis for the multi-region wind turbine controller can now be developed further.

\section*{Acknowledgment}

This research is part of the EU-Project POSYTYF
(POwering SYstem flexibiliTY in the Future through RES), https://posytyf-h2020.eu. 
The POSYTYF project has received funding from the European Union’s Horizon 2020 
research and innovation programme under grant agreement No 883985.


\bibliography{Lit_HS_2024_02_22}

\end{document}

%% file: def_RT_v9.tex
%
%



\newcommand{\trans}{^{\text{T}}}

\newcommand{\Cset}{{\mathbb{C}}}
\newcommand{\Dset}{{\mathbb{D}}}
\newcommand{\Iset}{{\mathbb{I}}}
\newcommand{\Rset}{{\mathbb{R}}}
\newcommand{\Sset}{{\mathbb{S}}}
\newcommand{\R}{{\mathbb{R}}}

\def\bs#1{\mathbf{#1}}

\def\Obs#1{\mathbf{#1}}
\def\B#1{\mathbf{#1}}

\def\Obsym#1{\boldsymbol{#1}}

\newcommand{\OAn}{\mathbf{A}}

\def\Oai{\Obs{a}_i}
\def\Oci{\Obs{c}_i}
\def\ab{\Obs{a}}
\def\bb{\Obs{b}}
\def\cb{\Obs{c}}
\def\db{\Obs{d}}
\def\eb{\Obs{e}}
\def\fb{\Obs{f}}
\def\gb{\Obs{g}}
\def\hb{\Obs{h}}
\def\ib{\Obs{i}}
\def\pb{\Obs{p}}
\def\qb{\Obs{q}}
\def\ub{\Obs{u}}
\def\rb{\Obs{r}}
\def\vb{\Obs{v}}
\def\wb{\Obs{w}}
\def\xb{\Obs{x}}
\def\xib{\Obs{\xi}}
\def\yb{\Obs{y}}
\def\zb{\Obs{z}}

\def\alphab{\boldsymbol{\alpha}}
\def\xib{\boldsymbol{\xi}}
\def\zetab{\boldsymbol{\zeta}}
\def\nub{\boldsymbol{\nu}}
\def\chib{\boldsymbol{\chi}}
\def\Deltab{\boldsymbol{\Delta}}
\def\thetab{\boldsymbol{\theta}}
\def\Gammab{\boldsymbol{\Gamma}}
\def\lambdab{\boldsymbol{\lambda}}

\def\Ac{\Obsym{\mathcal{A}}}
\def\Bc{\Obsym{\mathcal{B}}}
\def\Cc{\Obsym{\mathcal{C}}}
\def\Dc{\Obsym{\mathcal{D}}}
\def\Ec{\Obsym{\mathcal{E}}}
\def\Fc{\Obsym{\mathcal{F}}}
\def\Sc{\Obsym{\mathcal{S}}}

\def\Kc{{\mathcal{K}}}
\def\Lc{{\mathcal{L}}}

\def\Ab{\B{A}}
\def\Bb{\Obs{B}}
\def\Cb{\Obs{C}}
\def\Db{\Obs{D}}
\def\Fb{\Obs{F}}
\def\Eb{\Obs{E}}
\def\Gb{\Obs{G}}
\def\Hb{\Obs{H}} 
\def\Ib{\Obs{I}} 
\def\Kb{\Obs{K}} 
\def\Lb{\Obs{L}} 
\def\Mb{\Obs{M}} 
\def\Nb{\Obs{N}} 
\def\Ob{\Obs{O}} 
\def\Pb{\Obs{P}}
\def\Qb{\Obs{Q}}
\def\Rb{\Obs{R}}
\def\Tb{\Obs{T}}
\def\Ub{\Obs{U}} 
\def\Yb{\Obs{Y}}
\def\Xb{\Obs{X}}
\def\0b{\Obs{0}}

\def\O0{\Obs{0}}

\newcommand{\fmax}{\overline{f}}
\newcommand{\fmin}{\underline{f}}
\newcommand{\zmax}{\overline{z}}
\newcommand{\zmin}{\underline{z}}
\newcommand{\xmax}{\overline{x}}
\newcommand{\xmin}{\underline{x}}
\newcommand{\omegamax}{\overline{\omega}}
\newcommand{\omegamin}{\underline{\omega}}

\newcommand{\DS}{\displaystyle}
\newcommand{\TS}{\textstyle}
\newcommand{\SStyle}{\scriptstyle}
\newcommand{\SSStyle}{\scriptscriptstyle}

\newcommand{\Durch}[2]{\frac{\DS #1}{\DS #2}}


\newcommand{\DDt}{\Durch{\text{d}}{\text{dt}}}
\newtheorem{assumption}{Assumption} 



\newcommand{\Trans}{^{\text{T}}}
\newcommand{\ti}[1]{\tilde{#1}}


\newcommand{\yP}{\dot{y}}
\newcommand{\yPP}{\ddot{y}}
\newcommand{\zetaP}{\dot{\zeta}}
\newcommand{\zetaPP}{\ddot{\zeta}}
\newcommand{\thetaP}{\dot{\theta}}
\newcommand{\thetaPP}{\ddot{\theta}}
\newcommand{\qP}{\dot{q}}
\newcommand{\qPP}{\ddot{q}}
\newcommand{\xP}{\dot{x}}
\newcommand{\xPP}{\ddot{x}}
\newcommand{\betaP}{\dot{\beta}}

\newcommand{\xB}{\B{x}}
\newcommand{\yB}{\B{y}}
\newcommand{\zB}{\B{z}}
\newcommand{\kB}{\B{k}}
\newcommand{\uB}{\B{u}}
\newcommand{\dB}{\B{d}}
\newcommand{\eB}{\B{e}}
\newcommand{\fB}{\B{f}}

\newcommand{\xtil}{\tilde{\B{x}}}
\newcommand{\xtilP}{\dot{\tilde{\B{x}}}}
\newcommand{\ytil}{\tilde{\B{y}}}
\newcommand{\Vp}{\dot{V}}

\newcommand{\xPB}{\dot{\B{x}}}
\newcommand{\yPB}{\dot{\B{y}}}
\newcommand{\zPB}{\dot{\B{z}}}

\newcommand{\xhat}{\hat{x}}
\newcommand{\yhat}{\hat{y}}
\newcommand{\zhat}{\hat{z}}

\newcommand{\xhatB}{\hat{\B{x}}}
\newcommand{\yhatB}{\hat{\B{y}}}
\newcommand{\zhatB}{\hat{\B{z}}}

\newcommand{\xhatP}{\dot{\hat{x}}}
\newcommand{\yhatP}{\dot{\hat{y}}}

\newcommand{\xhatPB}{\dot{\hat{\B{x}}}}
\newcommand{\yhatPB}{\dot{\hat{\B{y}}}}

\newcommand{\ktil}{\tilde{\kB}}

\newcommand{\hiz}{h_i(\zB)}
\newcommand{\hizhat}{h_i\left(\hat{\zB}\right)}

\newcommand{\ytilde}{\tilde{\B{y}}}
\newcommand{\fahat}{\hat{\B{f}}_a}
\newcommand{\ey}{\B{e}_y}
\newcommand{\eytilde}{\tilde{\B{e}}_y}
\newcommand{\eytildeP}{\dot{\tilde{\B{e}}}_y}
\newcommand{\etildeP}{\dot{\tilde{\B{e}}}}
\newcommand{\Ptilde}{\tilde{\B{P}}_2}

\newcommand{\MC}[1]{\mathcal{#1}}

\newcommand{\AB}{\B{A}}
\newcommand{\BB}{\B{B}}
\newcommand{\CB}{\B{C}}
\newcommand{\DB}{\B{D}}
\newcommand{\EB}{\B{E}}
\newcommand{\FB}{\B{F}}
\newcommand{\IB}{\B{I}}
\newcommand{\LB}{\B{L}}
\newcommand{\NB}{\B{N}}
\newcommand{\PB}{\B{P}}
\newcommand{\TB}{\B{T}}

\newcommand{\Acal}{\MC{A}}
\newcommand{\Bcal}{\MC{B}}
\newcommand{\Ccal}{\MC{C}}
\newcommand{\Dcal}{\MC{D}}
\newcommand{\Ecal}{\MC{E}}

\newcommand{\AcalB}{\BS{\MC{A}}}
\newcommand{\BcalB}{\BS{\MC{B}}}
\newcommand{\CcalB}{\BS{\MC{C}}}
\newcommand{\DcalB}{\BS{\MC{D}}}
\newcommand{\EcalB}{\BS{\MC{E}}}
\newcommand{\FcalB}{\BS{\MC{F}}}

\newcommand{\AcalBz}[2]{\AcalB_{#1#2}\left(\B{z}\right)}
\newcommand{\AcalBzInv}[2]{\AcalB^{-1}_{#1#2}\left(\B{z}\right)}

\newcommand{\ABtil}{\B{\tilde{A}}}
\newcommand{\BBtil}{\B{\tilde{B}}}
\newcommand{\CBtil}{\B{\tilde{C}}}
\newcommand{\EBtil}{\B{\tilde{E}}}

\newcommand{\sumhiz}{\sum\limits_{i=1}^{N_r} h_i(\zB)}
\newcommand{\sumhiznolim}{\sum_{i=1}^{N_r} h_i(\zB)}
\newcommand{\sumhibeta}{\sum\limits_{i=1}^{N_r} h_i\left(\beta\right)}
\newcommand{\sumhibetanolim}{\sum_{i=1}^{N_r} h_i\left(\beta\right)}

\newcommand{\Atil}{\tilde{\B{A}}}
\newcommand{\Btil}{\tilde{\B{B}}}
\newcommand{\Etil}{\tilde{\B{E}}}
\newcommand{\AiCL}{\B{A}_{\text{CL},i}}
\newcommand{\AiCLtilde}{\tilde{\B{A}}_{\text{CL},i}}

\newcommand{\NormPtildeEy}{\left\|\Ptilde\,\eytilde\right\|}
\newcommand{\kappai}{\BS{\kappa}_i}
\newcommand{\normkappai}{\left\|\kappai\right\|}

\newcommand{\DandEVectorNoI}{\left[\DcalB_2 \: \EcalB_2 \right]}
\newcommand{\DandEVector}{\left[\DcalB_{2,i} \: \EcalB_{2,i}\right]}
\newcommand{\DandFVectorNoI}{\left[\DcalB_2 \: \FcalB_2 \right]}
\newcommand{\DandFVector}{\left[\DcalB_{2,i} \: \FcalB_{2,i}\right]}
\newcommand{\XiandFa}{\begin{pmatrix} \BS{\xi} \\ \B{f}_a\end{pmatrix}}
\newcommand{\XiHatandFaHat}{\begin{pmatrix} \hat{\BS{\xi}} \\ \fahat\end{pmatrix}}
\newcommand{\Kmax}{\MC{K}_\text{max}}
\newcommand{\nuB}{\BS{\nu}}
\newcommand{\rhoB}{\BS{\rho}}
\newcommand{\eps}{\varepsilon}

\newcommand{\VP}{\dot{V}}

\newcommand{\DTS}{\DcalB_2 \left(\B{z}\right)}
\newcommand{\ETS}{\EcalB_2 \left(\B{z}\right)}
\newcommand{\FTS}{\FcalB_2 \left(\B{z}\right)}

\newcommand{\rads}{\frac{\text{rad}}{\text{s}}}
\newcommand{\ms}{\frac{\text{m}}{\text{s}}}
\newcommand{\radssqr}{\frac{\text{rad}}{\text{s}^2}}

\newcommand{\om}{\omega}
\newcommand{\omrSP}{\omega_{r,\text{SP}}}
\newcommand{\omgSP}{\omega_{g,\text{SP}}}

\newcommand{\Fst}{F_\text{st}}

\newcommand{\p}{\partial}

\newcommand{\x}{\times}


\newcommand{\regsym}{\textsuperscript{\textregistered}}

\setlength{\parindent}{0cm}
\setlength{\parskip}{1ex}
